\documentclass[12pt,preprint]{aastex}

\def\rfe{R$_{\rm FeII}$}
\def\feiiq{\rm Fe{\sc ii }$\lambda$4570\/}

\def\ltsima{$\; \buildrel < \over \sim \;$}
\def\simlt{\lower.5ex\hbox{\ltsima}}            
\def\gtsima{$\; \buildrel > \over \sim \;$}

\def\simgt{\lower.5ex\hbox{\gtsima}}            

\def\ha{{\sc H}$\alpha$}

\def\civbc{{\sc{Civ}}$\lambda$1549$_{\rm BC}$\/}

\def\cm3{cm$^{-3}$\/}
\def\hb{{\sc{H}}$\beta$\/}
\def\hg{{\sc{H}}$\gamma$\/}
\def\hbbc{{\sc{H}}$\beta_{\rm BC}$\/}
\def\hbnc{{\sc{H}}$\beta_{\rm NC}$\/}

\def\oiiiopt{{\sc{[Oiii]}}$\lambda$4959,5007\/}
\def\o4363{{\sc{[Oiii]}}$\lambda$4363\/}

\def\feiiopt{{Fe \sc{ii}}$_{\rm opt}$\/}
\def\feii{{Fe\sc{ii}}$_{\rm opt}$\/}
\def\fe{{\sc{Fe}}\/}
\def\heii{{He}{\sc ii}$\lambda$4686\/}
\def\fe76087{{\sc [Fe vii]}$\lambda$6087\/}
\def\oiii{{\sc [Oiii]}$\lambda$5007}

\def\kms{km~s$^{-1}$}

\usepackage{graphicx}
\input epsf.sty
\usepackage{times}

\slugcomment{}

\shorttitle{Average QSO spectrum } \shortauthors{ Sulentic et al. }

\begin{document}


\title{Average Quasar Spectra in the Context of Eigenvector
1\altaffilmark{1}}


\author{J. W. Sulentic\altaffilmark{2 \dag}, P. Marziani
\altaffilmark{3 \dag}, R.  Zamanov\altaffilmark{3 \dag},  R.
Bachev\altaffilmark{2,3 \dag}, M. Calvani\altaffilmark{3 \dag}, D.
Dultzin-Hacyan\altaffilmark{4 \dag}}

\altaffiltext{1}{Based in part on data collected at ESO La Silla.}
\altaffiltext{2}{Department of Physics and Astronomy, University of
Alabama, Tuscaloosa, AL 35487, USA} \altaffiltext{3}{Osservatorio
Astronomico di Padova, Vicolo dell'Osservatorio 5, I-35122 Padova,
Italy} \altaffiltext{4}{Instituto de Astronom\'\i a, UNAM,
Apdo.Postal 70-264, 04510 Mexico D.F., Mexico} \footnotetext{$\dag$
``Northern Quasar Alliance''}


\begin{abstract}

Recent work has shown that it is possible to systematize quasar
spectral diversity in a parameter space called  ``Eigenvector 1''
(E1).  We present median AGN spectra for fixed regions of the E1
(optical) parameter space (FWHM(\hb) vs. equivalent width ratio
\rfe=W(\feiiq)/W(\hb). Comparison of the median spectra for different
regions show considerable differences. We suggest that an E1-driven
approach to median/average spectra emphasizes significant differences
between AGN, and offers more insights into AGN physics and dynamics
than a single population median/average derived from a large and
heterogeneous sample of sources. We find that the \hb\ broad component
line profile changes in shape along the E1 sequence both in average
centroid shift and asymmetry. While objects with FWHM(\hbbc)$\la$ 4000
\kms\ are well fitted by a Lorentz function, AGN with FWHM(\hbbc)$\ga$
4000 \kms\ are well fitted if two broad line components are used: a
broad (the ``classical'' broad line component) and a very
broad/redshifted component.
\end{abstract}


\section{Introduction}

Studies of broad emission line spectra for AGN provide both the
strongest constraint on models for the nebular physics and kinematic
models of the broad line emitting clouds.  Ideally one would like to
have both high signal-to-noise ratio (S/N) and moderate resolution
(\ltsima 5 \AA) measures for the strongest high and low ionization
lines as well as measures across as wide a wavelength range as possible
in order to better characterize the continuum shape.  The former
measures have led to the Eigenvector 1 parameter space concept
(Sulentic et al.  2000a,b) that is, in part, built upon foundations
laid almost ten years ago (Boroson \& Green 1992; hereafter BG92). The
above goals are very observing time intensive but recent AGN surveys
have provided an alternate method for examining quasar spectra from
1000 -- 10000 \AA. Average or composite spectra derived from large
survey databases involving hundreds of AGN (LBQS: Francis et al. 1991;
HST: Zheng et al. 1997; 1998; FIRST:  Brotherton et al. 2001; Sloan
survey: Vanden Berk et al. 2001) make it possible to generate a
``typical'' quasar spectrum from blueward of Ly$\alpha$\ to \ha.

What is unclear is whether such composite spectra are astrophysically
useful beyond, perhaps, allowing us to identify lines that are too weak
to be seen in individual source spectra.  The fundamental question is
whether the similarities or the differences in AGN line and continuum
phenomenology tell us more about the underlying physics. The
Eigenvector 1 concept has been advanced as a possible ``H-R Diagram"
for AGN in the sense that it appears to provide parameter space
discrimination between all major classes of broad line sources as well
as a correlation for, at least, radio quiet sources (Sulentic et al.
2000b).  The correlation and distribution of radio-quiet sources (with
FWHM(\hb)$\leq$ 4000 \kms) in E1 have been reasonably well fit with a
model that sees accretion rate, convolved with source orientation, as
the principal physical driver (Marziani et al.  2001). The generality
of E1, of course, requires much further testing but unless it is far
off the mark, it suggests that average or composite spectra should be
viewed in the same way that one would view an average stellar spectrum
taken over the full effective temperature range that is observed in the
H-R Diagram (spectral types O-M). Interpretation of composite spectra
from heterogeneous samples will be complicated in two ways: (1) they
will be  subject to selection biases dependent on the relative number
of sources with each ``spectra type'' in a given sample and, more
significantly, (2) they will average the spectra of sources with
dramatically different physical properties.  We suggest that the most
useful approach to averaging AGN spectra lies within the E1 context. We
present average spectra for fixed domain quadrants in E1 followed by a
brief discussion of important differences. The emphasis is on the
FWHM(\hb) measures although large differences in the equivalent width
parameter are also seen. Full discussion of the equivalent width
differences must wait for significant numbers of very high S/N measures
for \feii\ weak sources.

\section{Sample, Data Reduction and Analysis \label{anal}}

Our current sample includes optical spectra for 187 sources obtained
between 1988 and 1996.  They cover the wavelength range that includes
\hg, \hb, and most of the optical \feii\ emission (4200-5700 \AA\ in
the rest frame).  Spectra were obtained with the following telescopes
and spectrographs:  ESO 1.5m (B\&Ch), San Pedro Martir 2.2m (B\&Ch),
Calar Alto 2.2m (B\&Ch), KPNO 2.2m (Gold), and Mount Ekar (Asiago)
1.82m (B\&Ch). Data for 52 objects can be found in Marziani et al.
(1996) and the unpublished part of the dataset will appear in a
forthcoming paper (Marziani et al. 2002).  Spectra were taken with
very similar instrumental setups including typically 120\AA/mm
dispersion and 2 arcsec slit width yielding resolution in the range
4-7 \AA\ FWHM.  The  S/N  is typically in the range $\approx$ 20-30.
Spectra with S/N $\la$ 10 were excluded. The sample includes spectra
for 187 type 1 AGN with redshifts z $\la$ 0.8, $m_V\la$ 16.  Table 1
summarizes the properties of our data sample within the context of
Eigenvector 1 (see \S \ref{types} for bin definition). Composite
spectra were obtained by taking the median of all spectra with 1\AA
binning.  The S/N of the composite spectra was $\ga$ 100 except for
bin A3 where only 5 objects were available, and S/N$\approx$50. No
composite spectra were obtained for the rare objects in the outlier
domain (FWHM(\hbbc)$\ga$ 4000 \kms\ and \rfe $\ga$ 0.5) which spans
several E1 bins.

The following analysis procedures were applied to each spectrum (see
Marziani et al. 1996 for details). (1) Deredshifting using the narrow
line component of \hb, and/or \oiiiopt\ as rest frame measures. (2)
Normalization using the best estimate for a local continuum at
$\lambda\approx$ 5100 \AA. (3) Subtraction of \feiiq\ emission blends
using a template based on the spectrum of IZw1 which also yielded the
FeII blend equivalent width W(\feiiq). (4) Subtraction of the
\oiiiopt\ and \heii\ lines. (5) Subtraction of the narrow component
from \hb\ (\hbnc) using a Gaussian model. In cases where  no clear
inflection between broad and narrow components was seen, two
alternative approaches were employed: (a) we assumed that FWHM(\hbnc)
$\approx$\ FWHM(\oiii) or (b) we assumed that FWHM(\hbbc)
$\approx$\ FWHM(\feiiq). The latter approach was useful for Narrow Line
Seyfert 1 sources.  Following these procedures we extracted \hbbc\ and
derived equivalent width and FWHM measures which allowed us to populate
E1. Host galaxy contamination was not significant for the luminous AGN
included in this sample. The typical uncertainty for FWHM(\hbbc) is
about 10\%\, and for W(\hbbc) and W(\feiiq) about 10-15\%.
Errors on \rfe\ are estimated to be less than 0.2.

\section{Results}

\subsection{Spectral Types \label{types}}

\begin{figure}
\includegraphics[width=9.0cm]{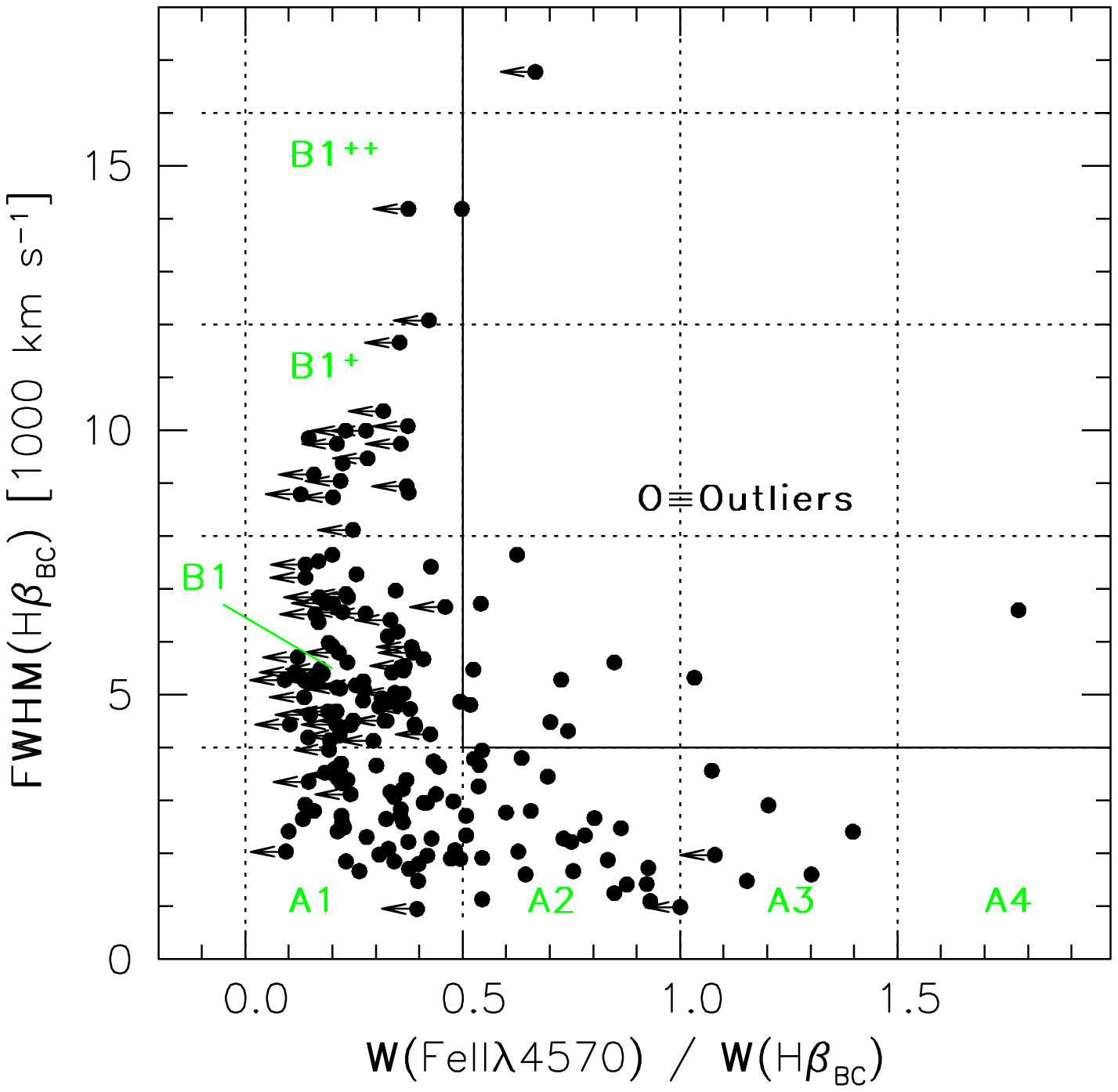}
\caption{ The  optical parameter plane (\rfe-FWHM(\hbbc)) of
Eigenvector 1. This is the largest sample that yet
displayed in an E1 context. Dotted lines correspond to the borders
between different spectral types also listed in Table 1. The abcissa
is \rfe=W(\feiiq)/W(\hbbc). The thin solid line indicates the
boundary of ``outlier'' sources and is, so far, a zone of avoidance
in quasar parameter space. Horizontal arrows denote \feiiq\ upper
limits.  Note that upper limits dominate single \rfe\ measurements
in the B1 and B1$^+$\ domains, while the higher S/N ratio of the median
spectra allow for a more accurate determination. } \label{FeHb}
\end{figure}

The distribution of 187 sources in the optical E1 plane (FWHM(\hbbc)
vs. \rfe) is shown in Figure \ref{FeHb}. We binned the optical
parameter plane as shown in Figure 1 and as defined in Table 1. The
bins were arbitrarily set to constant $\Delta$ FWHM(\hbbc) = 4000 \kms,
and $\Delta$ \rfe = 0.5.  A finer subdivision is not warranted at this
time by the accuracy of the measures or the heterogeneity of the data
sample.  The adopted A-B bin labeling reflects our earlier suggestion
that two distinct radio-quiet AGN populations may exist: Pop. A
(``pure'' radio-quiet) and Pop. B (same as the radio-loud domain; see
Table 1).  The numeral accompanying the letter designation for A bins
reflects the increasing strength of \rfe. The ``+'' designations for B
bins reflect increasing FWHM(\hbbc).  There are many measures that
distinguish between radio-quiet Pop. A and B as defined with a boundary
at FWHM(\hbbc) $\approx$ 4000 \kms. There are also many measures that
support a phenomenological commonality between population B and
radio-loud sources (Sulentic et al. 2000a,b,c).  Table 1 summarizes the
properties of domain space bins that are occupied by a significant
number of sources.  The composite (median) spectra after processing
step 2 (continuum normalized) and after step 3 (\feii\ subtracted) are
presented in the left and right panels of Figure 2 respectively. The
median spectra are available in digital format on the World Wide Web.
\footnotemark[6]\footnotetext[6]{At the address
http://panoramix.pd.astro.it/$\sim$marziani/medians.zip}

It is important to note that the E1 binned median spectra are unlikely
to be strongly influenced by any luminosity dependence since: (a) there
is,so far, no convincing evidence for a difference in line profile
properties with source redshift/luminosity (Sulentic et al. 2000a) and
(b) for any sub-sample, $\Delta M_B \la 1.5 \la \sigma_{M_B}$\ (see
Table 1). It is also worth noting that the median $M_B$\ of our sources
is close to the limit separating Seyfert 1 galaxies and quasars
($M_B$=-23.0). This should not be considered a particular problem
because there is no discontinuity in the AGN luminosity distribution.
Radio-loud sources in our sample show a higher mean luminosity perhaps
due to the inclusion of many beamed sources but Pop. B  radio-quiet
sources, co-spatial with them in E1, are not more luminous than Pop. A
radio-quiet sources.

Our  W(\feiiq) uncertainty estimate (\S \ref{anal}) is valid if
\feiiq\ is clearly detected above the noise. However, the minimum
W(\feiiq) for which it is possible to visually detect \feiiq\ depends
both on S/N and on \feiiq\ FWHM. For S/N $\approx $ 20. upper limits
for \feiiq\ detection is estimated to be $\approx$ 14, 17, 20\AA\ for
FWHM = 2500, 5000, 7500 \kms\ respectively. The  utility of the
composite spectra is also evident because of the much higher S/N in bins
B1 (and B1$^+$), which contain many upper limits because \feiiq\
is intrinsically weak in Population B sources. In those bins it is
possible to obtain more reliable W(\feiiq) measurements from composite
spectra.

\subsection{Median \hbbc\ Profiles Across E1}

Median spectra of the \hbbc\  profile are shown in Fig.  \ref{fits}.
They were extracted from the median spectra after \feiiopt\ and
continuum subtraction to take advantage of the much higher S/N of the
compositer spectra. We derived a pure \hbbc\ profile by first
subtracting \oiiiopt\ and \heii\ and then fitting \hbbc\ with a
high-order spline function  to avoid the effects of finite S/N and to
interpolate across the \hbnc. This approach has proven to be successful
for measuring line parameters in noisier line profiles (Marziani et
al.  1996).  It is preferable to template fitting because it makes no
assumptions about the shape of the profile.  Table 1 presents
FWHM(\hbbc) as well as line centroids at 1/2 and 1/4 maximum
intensities derived from the median spectra.  The median
\hbbc\ profiles in Pop. A are almost symmetric with a slight blueward
asymmetry in bin A2. The blue asymmetry becomes visually apparent in
bin A3 which contains sources that are extreme in many ways. Table 1
shows that the profiles of NLSy1 do not differ from bin A1 and A2
median spectra. The smaller median FWHM \hbbc\ reflects an arbitrary
cutoff ($\leq$2000 \kms) in the definition of NLSy1s.

The profiles of Pop. A sources are very different from those of Pop. B.
The median profiles of population B sources are red asymmetric with the
strongest asymmetry in the (few) bin B1$^{++}$ sources. The different
shape may point towards different BLR structure/kinematics in these
sources. Profile shape was not selected {\it a priori} in the
population definition that was based on a possible FWHM(\hbbc) break.
An alternate way to describe profile asymmetry involves fitting a
functional form to the median profiles. The \hbbc\ profile in NLSy1
sources is well fit by a Lorentz function (see also V{\'e}ron-Cetty,
V{\' e}ron, \& Gon{\c c}alves 2001). The same is generally true for  A1
and A2 median profiles while A3 \hbbc\ is based on only five sources.

\begin{figure*}[htb]
\includegraphics[width=16.0cm, height=9.0cm, angle=0]{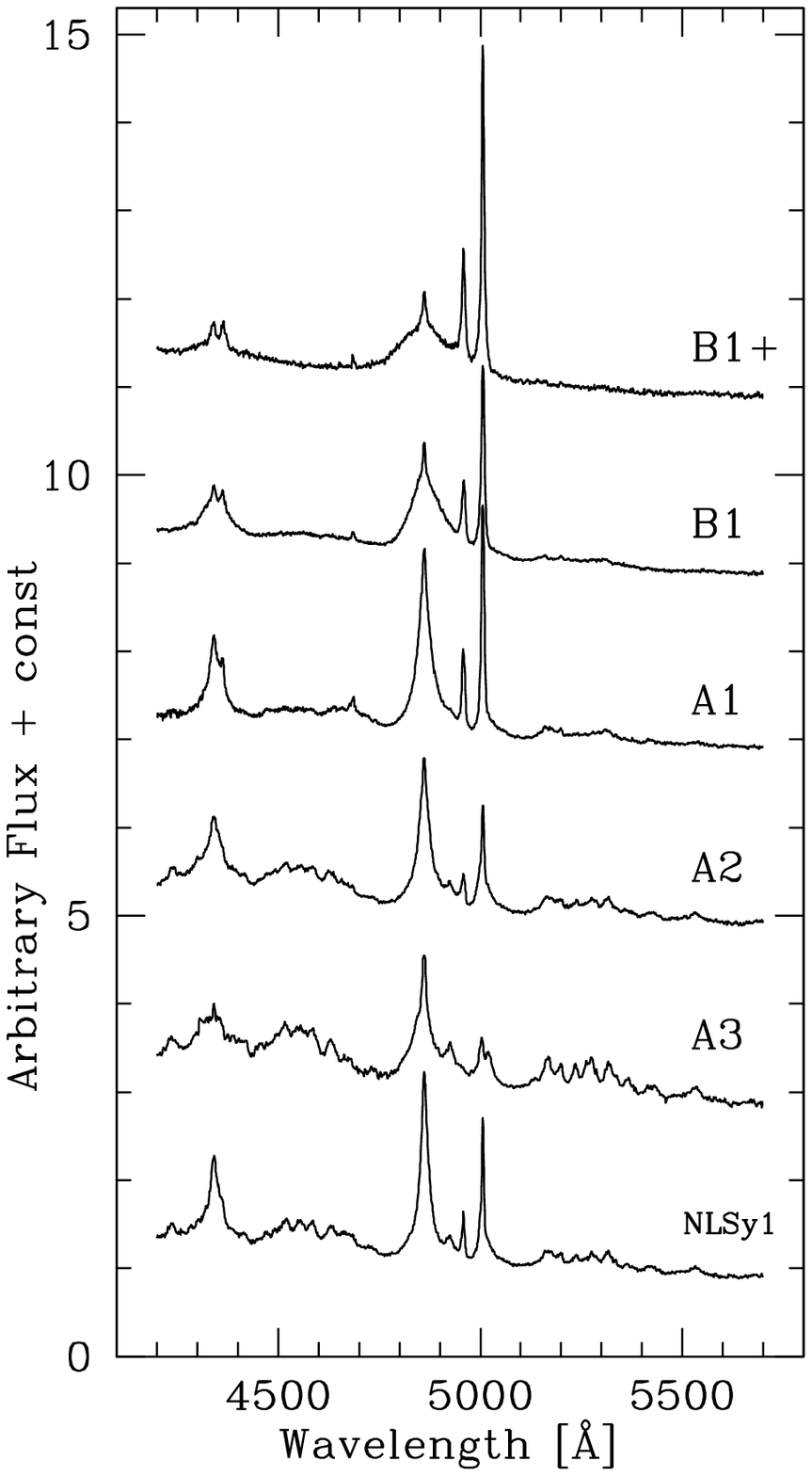}
\includegraphics[width=16.0cm, height=9.0cm, angle=0]{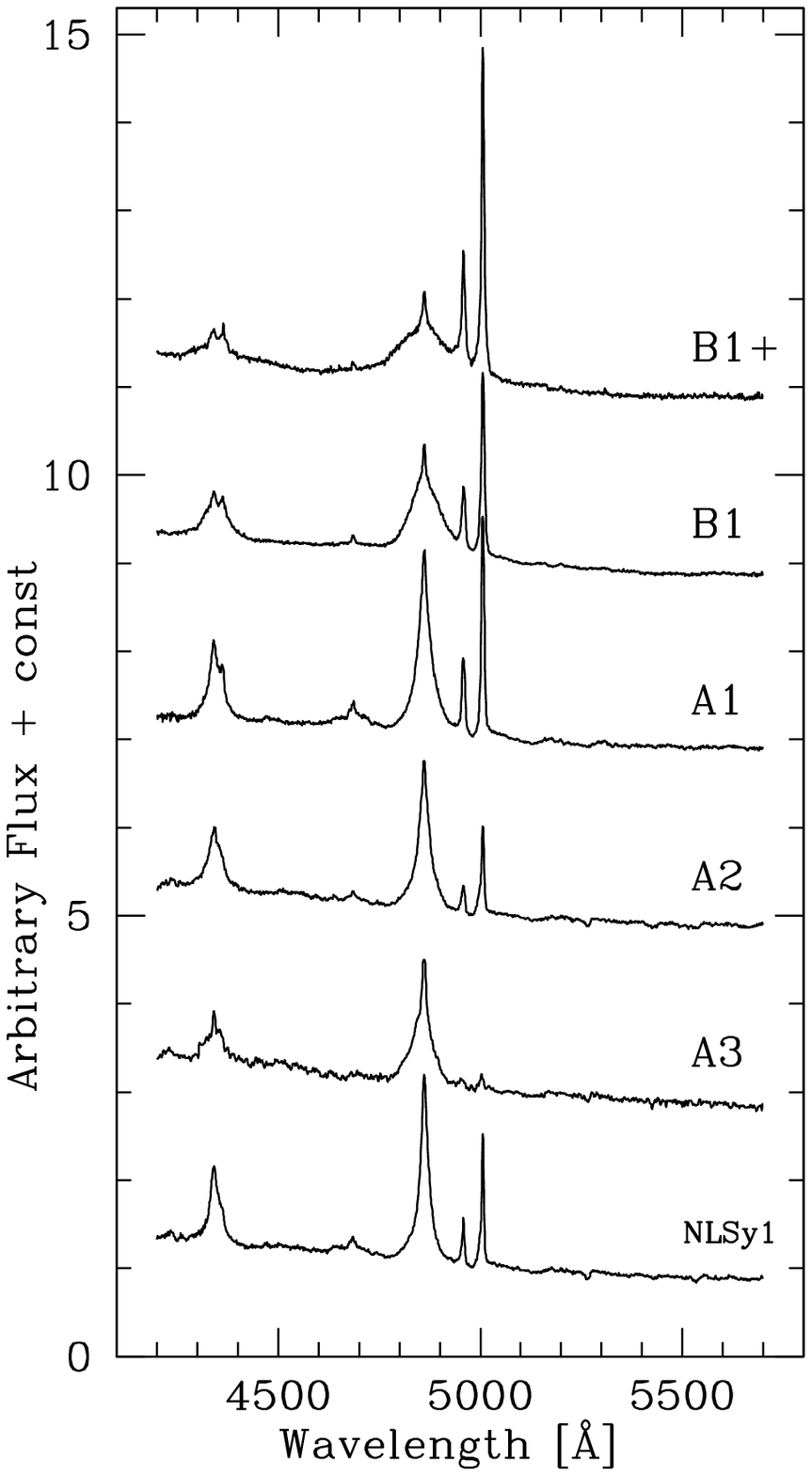}
\caption{Left panel: Composite  (median) spectra of the E1 parameter
bins defined in Table 1 and Figure 1.  Right panel: the same composite
spectra with \feii\ emission subtracted.} \label{Aver2}
\end{figure*}

\begin{figure*}[htb]
\includegraphics[width=9.0cm, height=8.0cm, angle=0]{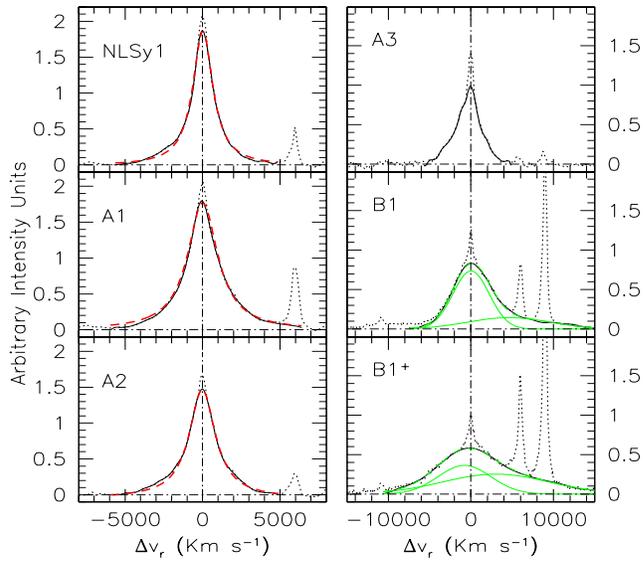}
\caption[]{\feii, and continuum subtracted \hb\ composite line profiles for the
different E1 parameter bins defined in Figure 1.  Solid
lines show the profile \hbbc\ after subtraction of narrow line
emission components. A Lorenztian fit (dashed line) is superimposed
on the NLS1, A1, A2 profiles. The individual components of a double
Gaussian (thin lines) and resultant fit (dashed line) are shown for
B1 and B1$^+$.} \label{fits}
\end{figure*}

A Lorentzian function is definitely not a satisfactory description
for bin B median profiles. The \hbbc\ profile is significantly
different and this can be seen even in Fig.  \ref{Aver2}. The
profile becomes more Gaussian but we were unable to fit bin B
profiles with any simple and physically meaningful function. The
best fits to \hbbc\ bin B1 and B1+ profiles required the sum of two
Gaussians: (1){\em  a FWHM$\approx$ 4000-5000 \kms  unshifted
Gaussian core} and  {\em (2) a broader FWHM$\approx$ 10000\kms
redshifted ($\Delta v_r \sim 1000 $ \kms) Gaussian base}.  The main
difference between B1 and B1$^+$ may be related to the strength of
the ``very broad component''. We will show in the next section: (1)
that many radio-loud and some radio-quiet Pop. B sources show
unambiguous very broad components and (2) that the interpretation of
the profile in terms of two components may have a straightforward
physical meaning.

\section{Discussion}

We find significant and systematic differences in median spectra across
E1. FWHM(\hbbc) and \rfe\ both show systematic changes and the other E1
parameters reinforce this dichotomy (pop. A sources show a systematic
blueshift of the CIV$\lambda$1549 broad line profile and and a soft
X-ray excess while pop. B source do not: Sulentic 2000ab).  Pop. A
sources show a more symmetric profile with the largest \rfe\ values
(indicating large W(\feiiq) and somewhat depressed W(\hbbc)). As one
enters the population B domain the radio-loud source fraction
increases.  \hbbc\  becomes increasingly broad and red asymmetric for
both radio-loud and radio-quiet sources. Overall NLSy1/Pop. A spectra
show lower ionization spectra than Pop. B (Marziani et al.  2001).
Comparison between high (\civbc: HIL) and low (\hbbc: LIL) ionization
emission lines suggests that they are emitted in disjoint regions in
pop. A.  Contrarily, both LIL and HIL may arise from a single region in
Pop. B sources (Marziani et al. 1996:  Sulentic et al. 2000a,b).  It is
more accurate to say that we cannot rule out the possibility that both
HIL and LIL arise from the same line emitting regions (broad and very
broad) in population B (Sulentic 2000a). The \hbbc\ profile analysis
indicates that the BLR -- and specifically the region emitting the low
ionization lines -- is most probably structurally different in
Population A and Population B sources. It is beyond the scope of the
present paper to report on detailed model calculation. We note however
that the Lorentz profile is consistent with emission from an extended
accretion disk.  This reinforces the suggestion that the LIL spectra in
Pop. A sources arise from a disk. The situation is less clear for Pop.
B sources where the Eddington ratio may be much lower. There is good
evidence that sometimes only one of the two emission components is
present in Pop. B sources (pure BLR or pure VBLR; see Sulentic et al.
2000c). In a companion paper, we suggest that even the Narrow Line
Region structure/kinematics may be changing along the E1 sequence
(Zamanov et al. 2002).

\subsection{The Very Broad Line Region Issue}

Can the double Gaussian model that is needed to fit Pop. B (and
radio-loud) profiles be physically justified? Several lines of
evidence point towards the existence of a Very Broad Line Region
(VBLR) at the inner edge of the BLR (Corbin 1997a,b).  Emission from
this region may be thought of as a sort of inner, large covering
factor, ``boundary layer'' where gas begins to become optically
thick (Marziani \& Sulentic 1993; Sulentic et al. 2000c). In objects
like the luminous quasar PG 1416-129, almost pure VBLR emission
survived after a continuum intensity decrease effectively quenched
the classical BLR. 3C232 (Marziani et al. 1996) may represent a good
radio-loud analog of PG1416. Other examples of possible pure VBLR
line sources can be found in Sulentic et al. 2000c. \heii\ has also
been known for some time to show a profile systematically broader
than  \hb\ in some sources (see Osterbrock \& Shuder 1982) and this
can be seen in Figure 2. Unfortunately \heii\ is usually weak and/or
blended with stronger \hb\ and FeII emission (it is on the red end
of \feiiq).  Detection of the \heii\ broad component is therefore a
function of \feii\ strength and S/N. The strongest very broad \heii\
feature is seen in the bin A1 median spectrum where we have  a
favorable combination of S/N, \hb\ width, \feii\ weakness and
\heii\ strength. The latter bias reflects the inclusion of data from
an observing run dedicated to studying strong \heii\ emission. A
very broad \heii\ profile is sometimes seen in pop. A sources (e.g.
Ton S 180) with much narrower \feiiq\.  However, in I Zw 1, the
highest S/N spectra fail to detect any trace of \heii\ emission. At
this time the question of \heii\ emission must be considered on an
object-by-object basis. We note that \heii\ strength was an
orthogonal (eigenvector 2) parameter in the principal component
analysis of Boroson \& Green (1992).

\section{Conclusion}

We generated average QSO spectra  in the 4200--5700 \AA\ spectral
region for fixed optical parameter bins in E1. In fixed parameter range
bins the mean spectral type appears to change systematically across
E1.  This systematic behaviour is reflected in  E1 X-ray and UV
measures as well. Composite spectra in an E1 context should provide
useful input for theoretical modeling. Ignoring the diversity of the
AGN \hbbc\ profile may make any theoretical modeling of the BLR
kinematics and dynamics unreliable.

\begin{acknowledgements}
The authors acknowledge  support from the Italian Ministry of
University and Scientific and Technological Research (MURST) through
grant  and Cofin 00$-$02$-$004.
\end{acknowledgements}

\clearpage


%
\clearpage
\begin{deluxetable}{cccccccccccc}
\tabletypesize{\scriptsize} \tablecolumns{12} \tablewidth{0pt}
\tablecaption{Sample  Properties }

\tablehead{ \colhead{Type} & \colhead{\rfe} & \colhead{FWHM(\hbbc)}
& \colhead{N$_{obj}$} &  \colhead{N$_{RL}$} &
\colhead{$M_B$\tablenotemark{a}} & \colhead{$\sigma_{M_B}$} &
\colhead{$\overline{\rm FWHM(H\beta_{BC})\tablenotemark{b}}$}
&\colhead{$\overline{\rm R_{\rm FeII}}$}
& \colhead{$\overline{c(1/2)\tablenotemark{b,c}}$} &
\colhead{$\overline{c(1/4)}$\tablenotemark{b,c}} & \colhead{Best Fit} \\
&  &   \colhead{[\kms]} & & & & & \colhead{[\kms]}& &
\colhead{[\kms]} & \colhead{[\kms]} &
\\
} \startdata
Total    &    all      &   all           & 187 & 64&  -23.7  & 2.0 &
\nodata &0.37 &  \nodata & \nodata & \nodata \\
NLSy1    &    all      &   $\leq$ 2000   &  24 &  2&  -23.0 &  1.7 & 1500
& 0.61& 40& 0  & Lorentz \\
Pop A\tablenotemark{d}    &    all      &   $\leq$ 4000   &  80 & 11&
-23.3 &  1.8 & 2300 & 0.49& 20 & 30 & Lorentz\\
A3       & 1.0 -- 1.5  &     0 --  4000  &   5 &  0&   -24.5 & 2.2 & 2350
&1.23& -280& -290 & Lorentz\\
A2       & 0.5 -- 1.0  &     0 --  4000  &  26 &  3&   23.3 &  1.5 & 1950
&0.70& -20 & -30 &Lorentz \\
A1       & 0.0 -- 0.5  &     0 --  4000  &  49 &  6&  -23.2 &  1.8 & 2250
&0.31& 30 & 250 & Lorentz \\
Pop B\tablenotemark{e}    &    all      &   $>$4000       &  97 & 50&
-24.2& 2.0 &  5700 &0.22& 200 & 700 & Double Gaussian\\
B1       & 0.0 -- 0.5  &  4000 --  8000  &  76 & 35  &   -24.2 & 2.0 &
5600 &0.23& 250 & 700 & Double Gaussian\\
B1$^+$       & 0.0 -- 0.5  &  8000 -- 12000  &  17& 12 &   -24.1 & 2.1 &
10000 &0.17 & 270& 1700 & Double Gaussian\\
B1$^{++}$       & 0.0 -- 0.5  & $>$12000        &   4& 3 & \nodata  &
\nodata & \nodata & \nodata & \nodata&\nodata &\nodata\\
outliers & $>$0.5      & $>$4000         &  10 & 3 &  -23.0  & 2.3 &
\nodata &0.80& \nodata& \nodata &\nodata\\
\enddata
\tablenotetext{a}{Absolute B magnitude from the V\'eron-Cetty and
V\'eron Catalogue, for $H_0=50$ \kms\ Mpc$^{-1}$, $q_0=0$.}
\tablenotetext{b}{Values measured on a high order spline function
representative of the \hbbc\ of each median spectrum.}
\tablenotetext{c}{Centroid at fractional intensity $\frac{i}{2}$ are
defined as c(i/4) = ($\lambda_B + \lambda_R - 2
\lambda_0)/\lambda_0$, where $\lambda_0 = 4861.33$ \AA, and
$\lambda_B$\ and $\lambda_R$\ are the wavelength of the blue and red
line side at the given fractional intensity.  }
\tablenotetext{d}{Pop. A $\equiv $ A1 $\bigcup$ A2 $\bigcup$ A3.}
\tablenotetext{e}{Pop. B $\equiv $B1$\bigcup$B1$^{+}
\bigcup$B1$^{++}$, with outliers (O) excluded}
\end{deluxetable}

\end{document}